\documentclass{article}

\usepackage{arxiv}

\usepackage[utf8]{inputenc} 
\usepackage[T1]{fontenc}    
\usepackage{hyperref}       
\usepackage{url}            
\usepackage{booktabs}       
\usepackage{amsfonts}       
\usepackage{nicefrac}       
\usepackage{microtype}      
\usepackage{lipsum}		
\usepackage{graphicx}
\usepackage{natbib}
\usepackage{doi}
\usepackage{longtable}
\usepackage{lscape} 
\usepackage{array}
\usepackage{authblk}
\usepackage{amssymb}
\usepackage{pifont}
\newcommand{\xmark}{\ding{55}}%

\title{Machine learning approaches for COVID-19 detection from chest X-ray imaging: A Systematic Review}

\author[1]{Harold Brayan Arteaga-Arteaga}
\author[1]{Melissa delaPava}
\author[1]{Alejandro Mora-Rubio} 
\author[1]{Mario Alejandro Bravo-Ortíz} 
\author[1]{Jesus Alejandro Alzate-Grisales} 
\author[1]{Daniel Arias-Garzón}
\author[2]{Luis Humberto López-Murillo}
\author[3]{Felipe Buitrago-Carmona}
\author[1]{Juan Pablo Villa-Pulgarín} 
\author[1]{Esteban Mercado-Ruiz} 
\author[3, 4]{Simon Orozco-Arias} 
\author[5]{M. Hassaballah}
\author[6]{Maria de la Iglesia-Vaya} 
\author[1]{Oscar Cardona-Morales}
\author[1,\thanks{Corresponding author: Reinel Tabares-Soto. Email address: rtabares@autonoma.edu.co}]{Reinel Tabares-Soto}

\affil[1]{Department of Electronics and Automation, Universidad Autónoma de Manizales, Manizales, Colombia}
\affil[2]{Department of Chemical Engineering, Universidad Nacional de Colombia, Manizales, Colombia}
\affil[3]{Department of Computer Science, Universidad Autónoma de Manizales, Manizales, Colombia}
\affil[4]{Department of Systems and informatics, Universidad de Caldas, Manizales, Colombia}
\affil[5]{Faculty of Computers and Information, South Valley University, Qena, Egypt}
\affil[6]{Unidad Mixta de Imagen Biomédica FISABIO-CIPF, Fundación para el Fomento de la Investigación Sanitario y Biomédica de la Comunidad Valenciana, Valencia, Spain}

\renewcommand{\shorttitle}{A Systematic Review}

\hypersetup{
pdftitle={A template for the arxiv style},
pdfsubject={q-bio.NC, q-bio.QM},
pdfauthor={David S.~Hippocampus, Elias D.~Striatum},
pdfkeywords={First keyword, Second keyword, More},
}

\begin{document}
\maketitle

\begin{abstract}
There is a necessity to develop affordable, and reliable diagnostic tools, which allow containing the COVID-19 spreading. Machine Learning (ML) algorithms have been proposed to design support decision-making systems to assess chest X-ray images, which have proven to be useful to detect and evaluate disease progression. Many research articles are published around this subject, which makes it difficult to identify the best approaches for future work. This paper presents a systematic review of ML applied to COVID-19 detection using chest X-ray images, aiming to offer a baseline for researchers in terms of methods, architectures, databases, and current limitations.
\end{abstract}


\section{Introduction}

The first confirmed cases of COVID-19 disease appeared in Wuhan, Hubei province, China back in December 2019 \citep{Wang2020AConcern.}. The disease is caused by the severe acute respiratory syndrome coronavirus 2 (SARS-CoV-2) \citep{Yuen2020SARS}, which is transmitted primarily through droplets of saliva or discharge from the nose \citep{WHOcoronavirus}. It has spread all over the world, and it was declared as a pandemic by the World Health Organization (WHO) in March 2020 \citep{WHOpandemic}. Based on data from Johns Hopkins University, as of October 13th, 2021, there have been 239,038,163 confirmed cases of COVID-19 around the world including 4,871,163 deaths \citep{WorldHealthOrganization2021}. According to the WHO, most infected people develop a moderate illness with symptoms such as fever, dry cough, and fatigue. In severe cases where patients need hospitalization, the symptoms include breathing difficulties, chest pain, and loss of speech or movement \citep{WHOcoronavirus}. 

COVID-19 can be diagnosed using tools based on the detection of viral gene, human antibody, or viral antigen, which requires qualified personnel and a specialized laboratory. As a complementary diagnostic tool, doctors employ medical imaging techniques such as chest X-ray or chest Computerized Tomography (CT). The produced images offer information about the lungs and can help radiologists to detect diseases like pneumonia, tuberculosis, interstitial lung disease, pneumothorax, early lung cancer, among others \citep{Anis2020AnRadiograph}. These images also have proven to be effective for COVID-19 detection, as well as giving information about disease progression through the evaluation of radiological findings \citep{Ng2020}.

\begin{itemize}
    \item \textit{Real-Time Reverse Transcription Polymerase Chain Reaction}: the viral gene detection by Real-Time Reverse Transcription Polymerase Chain Reaction (RT-PCR) is very sensitive because it can detect a copy of a specific genomic sequence, which has lead to the development of many commercial technologies that use nasal or nasopharyngeal swabs along with RT-PCR for COVID-19 detection \citep{Yuce2021COVID}.
    
    However, the detection of COVID-19 using RT-PCR is complex, the materials are sometimes slow to deliver, and the complete process can only be performed by qualified clinical laboratory personnel, which take over 24 hours from taking the sample to getting the analysis results. It is also expensive due to one kit can cost over 100 USD and setting a lab costs more than 15,000 USD. Additionally, many factors like storage, collection, processing, and genomic mutations, can lead to incorrect results \citep{Afzal2020}. Other drawbacks include low availability in some countries \citep{Aziz2020}, and high false-negative rates \citep{Fan2020Inf-Net:Images}. 
    
    \item \textit{Computed Tomography scan}: this is an advanced technique that allows to generates detailed 3D images of organs and soft tissues \citep{Ohata2021}. Unlike RT-PCR, a CT scan is fast to obtain and it is relatively easy to perform. It has been recently reported that this technique shows typical features of COVID-19 like ground-glass opacities and multifocal patchy consolidation, even in patients with negative PCR but clinical symptoms \citep{Ai2020}. However, decontaminate CT equipment after scanning COVID-19 patients may damage it. Thus, to minimize the risk of cross-infection, it is suggested to use portable devices like chest radiography, which is already a triage tool in many hospitals \cite{Wong2020}.
    
    \item \textit{Chest X-ray}: X-ray refers to a medical imaging technique that uses radiation to generate an image of internal structures of the human body. The main elements that are assessed using X-ray are bones, which appear white on the image; soft tissues, which appear as light gray; fat, which appears gray; and gas, which appears black \citep{Anis2020AnRadiograph}. In particular, a chest X-ray image allows a doctor to evaluate multiple organs, structures and conditions \cite{Breiding2014}. It is one of the most used methods to diagnose pneumonia worldwide \citep{JAISWAL2019511}. 

    Chest X-ray devices can be portable, are affordable, fast, and gives the patient a lower radiation dose than CT. It has been reported that common CT findings can also be detected on chest X-ray images, even in patients with initial negative RT-PCR for COVID-19. However, the diagnosis of COVID-19 using chest X-ray images is more difficult than using CT or other imaging modalities and can only be performed by specialist physicians, which scarce \citep{narin2021, Wong2020}.
\end{itemize}

There is a trade-off between quality and accessibility when choosing the imaging technique to use. CT produces a higher quality image but requires a much more complex device, not always available in many institutions. On the contrary, X-ray devices are much more affordable, can be portable, and are less harmful, given that a single CT scan can deliver a median effective radiation dose as high as 442 chest X-ray series \citep{Breiding2014}.
    
\begin{figure*}[!ht]
\centering
\includegraphics[width=\textwidth]{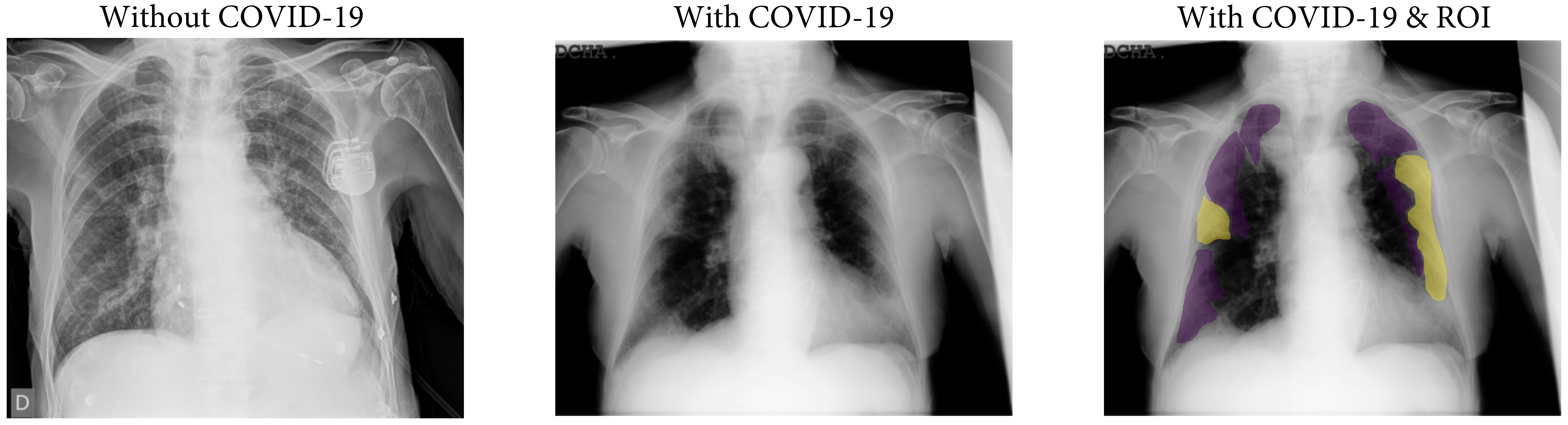}
\caption{Examples of chest X-ray images for a COVID-19 negative subject, a COVID-19 positive subject. The last one shows the masks over the regions of ground glass opacities (yellow) and consolidations (purple) on the region of interest.}
\label{fig:chestXRay}
\end{figure*}

Concerning the radiological findings on chest X-ray and CT associated with COVID-19 pneumonia, the most common are ground glass opacities. These marks are usually bilateral, meaning they affect both lungs and are more likely to be located in the periphery and lower areas of the lungs \citep{Kaufman2020Review, Yasin2020Chest}. They can be seen in chest X-ray images, or CT images as regions of increased whiteness due to the augmented density \citep{Cleverly2020TheRole}, which do not cover blood vessels and airway walls completely \citep{Rousan2020Chest}. As the disease progresses, this finding becomes denser and covers blood vessels and airway walls on the image, becoming consolidations. Fig. \ref{fig:chestXRay} presents a comparison of the chest X-ray images for a COVID-19 negative subject and a COVID-19 positive subject; additionally, for the COVID-19 case, the image on the right shows the masks over the regions of ground glass opacities (yellow) and consolidations (purple).

The WHO solidarity consortium from February 2021 presents a study to find a drug against COVID-19, however, it was found that the mortality, initiation of ventilation, and hospitalization duration were not definitely reduced by any trial drug. Until now, no specific drug has been found against COVID-19 \cite{WHOso2021}. Currently, there are 153 vaccine candidates, 476 vaccine trials ongoing, 23 vaccines approved by at least one country, and 7 vaccines approved for use by the WHO against COVID-19. The AstraZeneca vaccine is the one approved in the largest number of countries, followed by Pfizer/BioNTech, Moderna, and Janssen \citep{vaccines}.

Artificial Intelligence (AI) techniques, including Machine Learning (ML), can be used for COVID-19 diagnosis from chest X-ray images and set foundations for automatic decision-making support systems \citep{Arteaga-Arteaga2022DeepImages}. AI refers to the process of providing computer features from human intelligence. ML is a subset of AI that holds the mathematical models used to achieve this task, whereas Deep Learning (DL) is a subset of ML itself and relates the models and algorithms based on neural networks \citep{Goodfellow-et-al-2016}. In general, ML and DL techniques are designed to extract features and find relationships between data samples. Thereby, these approaches are well-suited for tasks relying on the human experience \citep{Orozco-Arias2019, Tabares-Soto2019, BravoOrtiz2021, 9328287, Arteaga-Arteaga2021MachinePatterns} such as classifying a chest X-ray image as positive or negative for COVID-19. Besides decision-making support systems in the medicine and healthcare field, AI has been used to perform tasks from managing medical data and analyzing health plans to drug development and health monitoring \citep{Amisha2019Overview}. AI applications in medicine aim to improve diagnostic performance and offer a better quality of service \citep{Ahuja2019}.

Regarding the detection of COVID-19 in chest X-ray images using AI techniques, most research papers propose a transfer learning approach using Convolutional Neural Networks (CNNs) such as VGG19, Inception, and MobileNet \citep{Pham2021ClassificationTuning}. A different approach creates novel CNNs to classify chest X-ray images as positive or negative for COVID-19 \citep{Hussain2021}. Different traditional ML approaches have been proposed involving a manual feature extraction stage employing texture or morphological descriptors of the images \citep{Hussain2020Machine-learningInfection, Pereira2020}. The availability of data is probably the biggest limitation when designing AI systems to detect COVID-19, although nowadays there are several public image databases, the quality of images and information is highly variable, which makes it difficult for researchers to evaluate their systems on appropriate conditions. Furthermore, there is not a standard benchmark to evaluate and compare the different proposals, which in combination with data variability, makes the reported results difficult to compare with each other.

The paper is organized as follows: Section Survey Methodology explains the criteria used to perform the literature review; Section Development Of The Subject presents the results and the state of the art; and Section Conclusions and Future Work.

\section{Survey Methodology} 
A systematic review of scientific papers was conducted, which explains the design and implementation of CNN, ML algorithms, or segmentation methods, for COVID-19 classification from chest X-ray images.

\subsection{Identification of the need for a review}
Given the need to develop more efficient and effective diagnostics tools for the COVID-19 disease, many research papers have been written since the beginning of the pandemic. One estimate suggests that more than 200,000 research papers have been published in journals and preprints repositories only in 2020 \citep{Else2020CovidPublishing}. Therefore there is a need for a new state-of-the-art review. 

State-of-the-art works up to March 21, 2021, are summarized in Table \ref{tab:related_works}, where is shown if they specify the corresponding preprocessing techniques and datasets used in the selected works, the date of search, whether or not it is systematic, the number of search databases used, the data modalities included in the work, the number of X-ray related articles analyzed and the number of X-ray related databases described. Table \ref{tab:related_works} shows that some of the available state-of-the-art works do not follow a systematic approach, and only include works up to July 2020 or before, unlike our work that covers from January 1, 2020, to March 21, 2021. Additionally, our work uses the largest number of search databases and some of the available state-of-the-art include a limited amount of papers that identify COVID-19 using X-ray images. 

The description of the available datasets presented in this work also fills a gap in the literature, since the available works barely describe some of them. We also present the preprocessing techniques and the specific datasets used on each work selected for this literature review, what is not included in any of the available works, along with the models, tasks and results described help researchers to build a complete panorama of the actual strategies used for the detection of COVID-19 using X-ray images. Our work is a complement to many of the available works since they invest more effort in discussing the risk of bias, recommendations, and deficiencies than in the specific methodologies details as we do.

Therefore, our bibliographic review contributes relevant and up-to-date information about the development of AI-based systems to detect COVID-19 from chest X-ray images. It will also offer a baseline for researchers regarding methods, architectures, databases, and current limitations.

\subsection{Research questions}
In order to describe the state-of-the-art approaches for COVID-19 detection, this paper aims to answer the following questions:
\begin{itemize}
 \item Which are the different architectures and novel components of CNNs used to detect COVID-19 on chest X-ray images?
 \item What are the detection performances of COVID-19 on chest X-ray images using CNNs?
 \item Which digital image databases are the most used for COVID-19 detection on chest X-ray images?
 \item Which segmentation methods are applied on chest X-ray images for the automatic detection of COVID-19?
\end{itemize}

\subsection{Bibliographic search} 
The key words chosen for this search are:
\begin{itemize}
    \item COVID-19.
    \item Deep Learning.
    \item Machine Learning.
    \item Classification.
    \item Segmentation.
    \item Chest X-ray.
\end{itemize}

After defining the search terms, the search string was built with logical operators. Due to COVID-19 being a disease that emerged in late 2019, the search is limited between 2020 - Present, only in the English language. The general search string is: \textit{Covid-19 AND (Machine Learning OR Deep Learning) AND (Classification OR Segmentation) AND X-ray}. Table \ref{Table:strings} shows the databases and search strings used for the review. The gray literature search included papers with novel COVID-19 classification methods from chest X-ray images.


\begin{landscape}
\begin{table}[!ht]
\caption{Related works}
\resizebox{1.4\textheight}{!}
{ 
\begin{tabular}{p{2.2cm}>{\centering\arraybackslash}p{2.5cm}>{\centering\arraybackslash}p{2.7cm}>{\centering\arraybackslash}p{2.3cm}>{\centering\arraybackslash}p{2.6cm}>{\centering\arraybackslash}p{2cm}>{\centering\arraybackslash}p{2cm}>{\centering\arraybackslash}p{2.4cm}>{\centering\arraybackslash}p{1.8cm}>{\centering\arraybackslash}p{2.8cm}}
\hline
\textbf{Article}     & \textbf{Database specification} & \textbf{Preprocessing specification} & \textbf{Date of search} & \textbf{Date of publication} & \textbf{Systematic} & \textbf{Search databases} & \textbf{Articles analyzed} & \textbf{Databases described}\\
\hline
\cite{Rahman2021a}       & \checkmark & \xmark   & -               & March 02, 2021       & \xmark   & -   & 23   & 12  \\
\cite{Nayak2021a}        & \xmark  & \xmark   & -               & March 20, 2021       & \checkmark  & 4   & 41   &  1  \\
\citep{Wynants2020}      & \xmark  & \xmark   & July 01, 2020   & April 07, 2021       & \checkmark  & 5   & 22   & -   \\
\cite{Shi2021}           & \xmark  & \xmark   & March 31, 2020  & April 16, 2020       & \xmark   & -   & 4    & -   \\
\cite{Swapnarekha2020}   & \xmark  & \xmark   & May 03, 2020    & May 26, 2020         & \xmark   & -   & 12   & -   \\
\cite{Salehi2020}        & \xmark  & \xmark   & -               & June 23, 2020        & \xmark   & -   & 5    & -   \\
\cite{Albahri2020}       & \xmark  & \xmark   & May 15, 2020    & June 25, 2020        & \checkmark  & 4   & 11   & -   \\
\cite{Bansal2020}        & \xmark  & \xmark   & -               & August 01, 2020      & \xmark   & -   & -    & -   \\
\cite{Syeda2021}         & \xmark  & \xmark   & June 27, 2020   & January 11, 2021     & \checkmark  & 3   & 22   & 23  \\ 
\cite{Roberts2021}       & \xmark  & \xmark   & -               & March 15, 2021        & \checkmark  & -   & 22   & -   \\  
Our work                 & \checkmark & \checkmark  & March 21, 2021  & -                    & \checkmark  & 6   & 23   & 18  \\
\hline
\end{tabular}
}
\label{tab:related_works}
\end{table}
\end{landscape}


\begin{table}[!ht]
\caption{\textit{Databases and search strings for literature review}}
\label{Table:strings}
\centering
\begin{tabular}{|c|l|}
\hline
\textbf{\begin{tabular}[c]{@{}c@{}}Name of the \\ search database\end{tabular}} & \multicolumn{1}{c|}{\textbf{Search string}}                                                                                                                          \\ \hline
Scopus                                                                              & \begin{tabular}[c]{@{}l@{}}TITLE-ABS-KEY (Covid-19 AND (Machine Learning OR Deep Learning) \\ AND (Classification OR Segmentation) AND   X-ray)\end{tabular}  \\ \hline
Web of Science                                                                      & \begin{tabular}[c]{@{}l@{}}(SUBJECT OR TITLE: Covid-19 AND (Machine Learning OR Deep Learning) \\ AND (Classification OR Segmentation) AND   X-ray)\end{tabular} \\ \hline
SpringerLink                                                                        & \begin{tabular}[c]{@{}l@{}}Covid-19 AND (Machine Learning OR Deep Learning) \\ AND (Classification OR Segmentation) AND X-ray.\end{tabular}                        \\ \hline
PubMed                                                                              & \begin{tabular}[c]{@{}l@{}}Covid-19 AND (Machine Learning OR Deep Learning)\\  AND (Classification OR Segmentation) AND X-ray.\end{tabular}                        \\ \hline
IEEE Xplore                                                                         & \begin{tabular}[c]{@{}l@{}}Covid-19 AND (Machine Learning OR Deep Learning) \\ AND (Classification OR Segmentation) AND X-ray\end{tabular}                         \\ \hline
Google Scholar                                                                      & \begin{tabular}[c]{@{}l@{}}Covid-19 AND (Machine Learning OR Deep Learning)\\  AND (Classification OR Segmentation) AND X-ray.\end{tabular}                        \\ \hline
\end{tabular}
\end{table}


\subsection{Inclusion and exclusion criteria}
The inclusion criteria taken into account are:
\begin{itemize}
    \item Papers published in Journals.
    \item Papers written in English.
    \item Papers found in the databases in Table \ref{Table:strings}.
    \item Papers that use DL o ML to detect COVID-19 from chest X-ray images.
    \item Papers using novel methods for COVID-19 detection.
\end{itemize}
    
The exclusion criteria taken into account are:
\begin{itemize}
    \item COVID-19 classification or segmentation papers without application of DL or ML methods.
    \item COVID-19 classification papers that do not use chest X-ray.
    \item Papers with methods that include X-ray and CT simultaneously in the methods training.
\end{itemize}

\subsection{Data extraction and synthesis}
\begin{figure*}[!ht]
    \centering
    \includegraphics[width=90mm, scale=0.2]{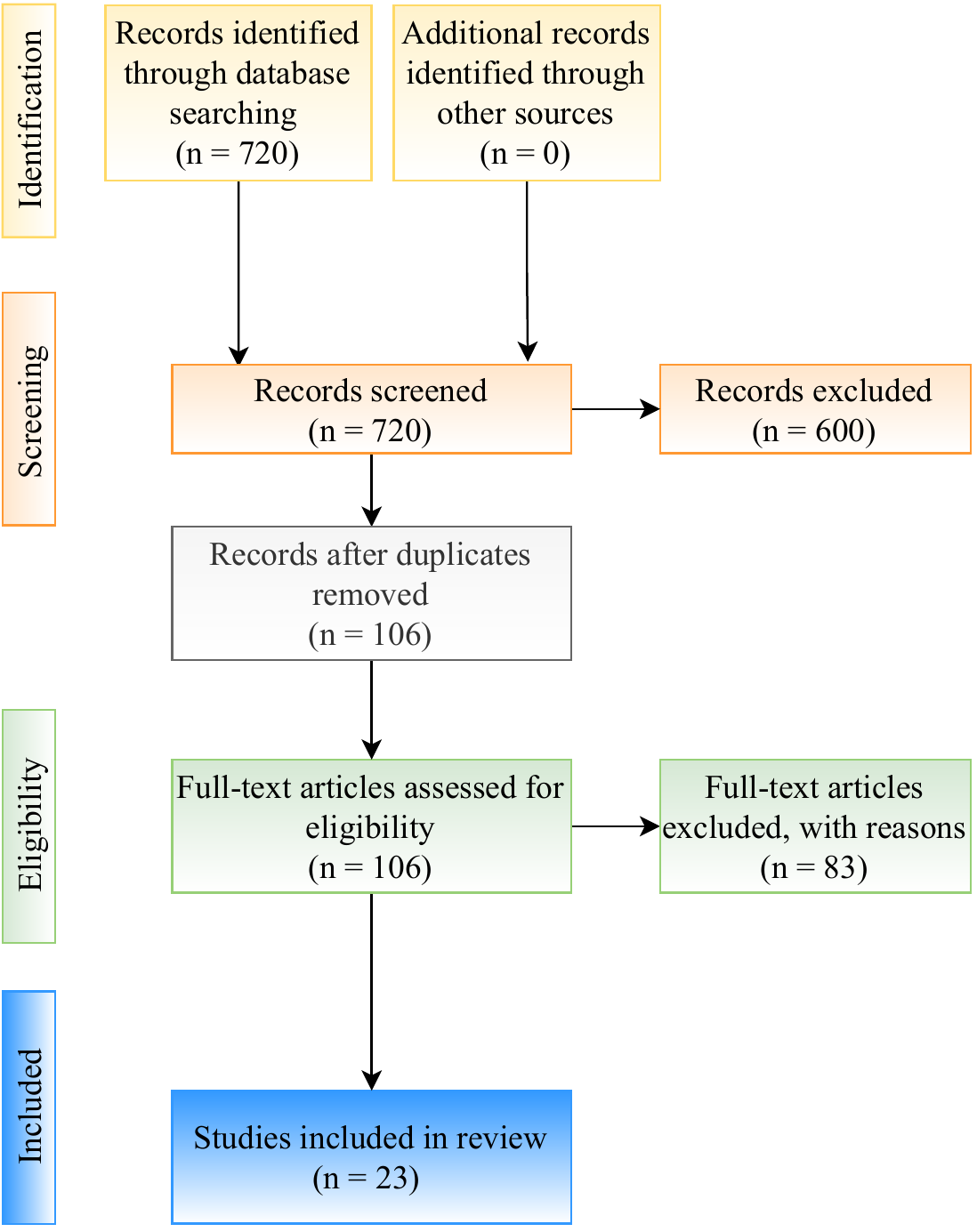}
    \caption{PRISMA flow diagram. PRISMA flow chart for search and article screening process. From: \citep{Moher2009PreferredStatement}}
    \label{fig:prisma}
\end{figure*}

We conduct a systematic literature review by applying the preferred reporting items for systematic reviews and meta-analyses guidelines (the PRIMA statement), the Fig. \ref{fig:prisma} presents the number of articles obtained on each step of the guidelines \citep{Moher2009PreferredStatement}.

Based on the search string provided in Table \ref{Table:strings}, approximately 720 papers are found and 120 papers per database are analyzed, from which 20 relevant papers per database are pre-selected. The pre-selection is done using the number of citations and the novelty of the COVID-19 detection method proposed. It is relevant to clarify that we found 14 repeated papers that are rejected; therefore, the pre-selection of the papers leads to 106 documents. 
The remaining 106 articles are filtered according to:

\begin{itemize}
    \item \textbf{Title:} After reading the title, 40 are accepted and 66 papers are rejected.
    \item \textbf{Abstract}: After reading the abstract, 28 are accepted and 12 papers are rejected.
    \item \textbf{Full text}: After reading the entire text, 23 papers are accepted and 5 papers are rejected.
\end{itemize}

\begin{figure*}[!ht]
    \centering
    \includegraphics[width=0.75\textwidth]{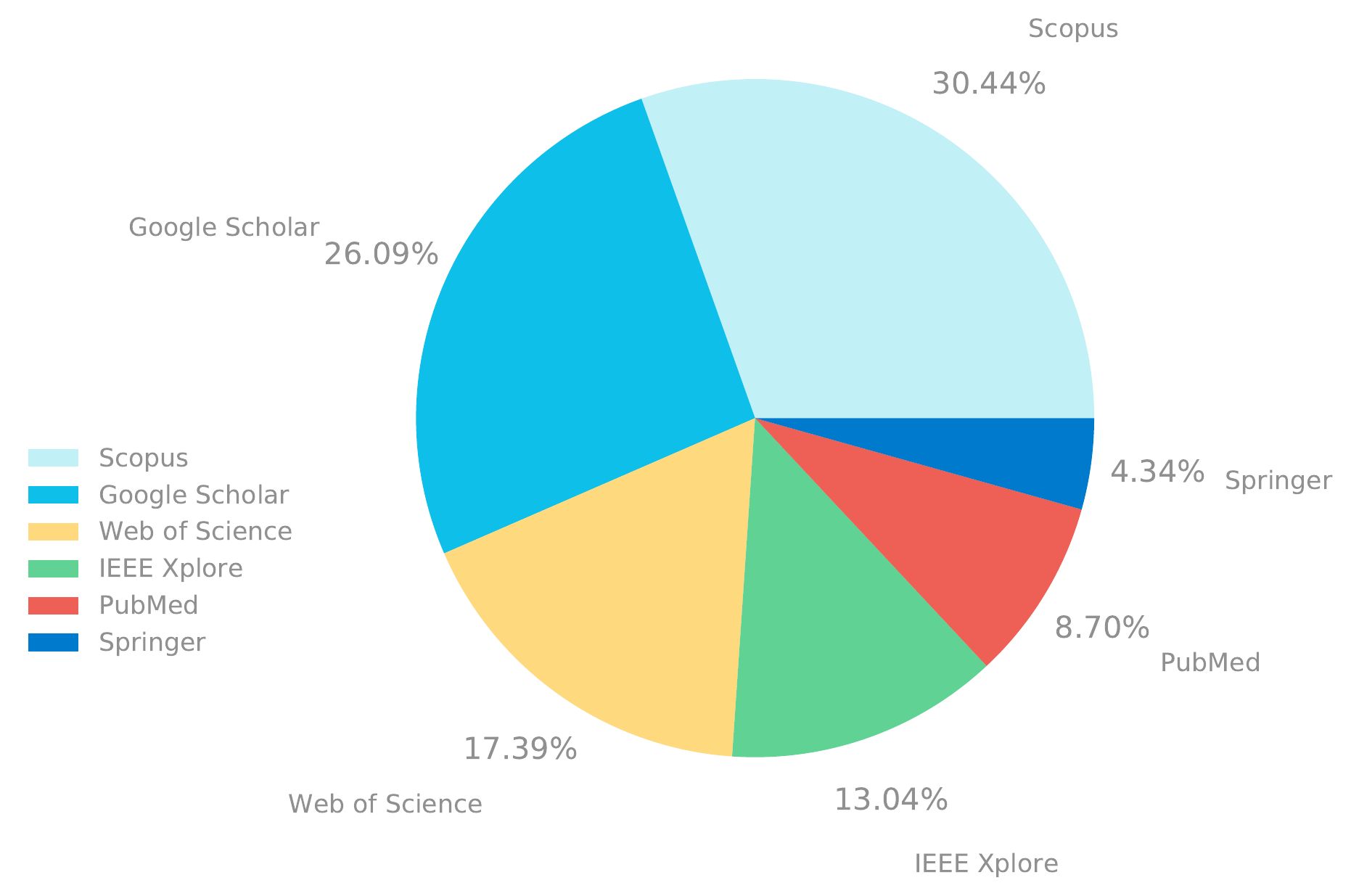}
    \caption{Percentage of papers selected for bibliographic review in different databases.}
    \label{fig:cake}
\end{figure*}

Figure \ref{fig:cake} presents the percentage distribution of the selected papers in the databases.

\section{Development of the subject}
As mentioned in the previous section, a total of 23 research papers are selected for the systematic review. This paper aims to cover novel approaches and offer a general overview of how AI has been applied to COVID-19 diagnosis.

\subsection{X-ray images databases}
Table \ref{table:databases} presents the most relevant databases used in the selected articles, they are described along with the total number of images, and the classes. This table includes not only databases with COVID-19 samples but also other chest diseases, due to some of the works published regarding the detection of COVID-19, which intend to classify multiple chest conditions. The most often classified illnesses along with COVID-19 are pneumonia viral, pneumonia bacterial, and tuberculosis. The COVID-19 Image Data Collection published by \cite{PaulCohen2020} is the most used database, it was one of the first ones to be released and more images are included over time. It also provides prospective metadata like survival, ICU stay, intubation events, blood tests, location, and freeform clinical notes. The most regularly used databases in this regard with X-ray samples of other chest diseases are ChestX-ray8 (CRX8), CheXpert, Chest X-ray Images (Pneumonia), and Tuberculosis chest X-ray.

\begin{longtable}[!ht]{|p{0.8cm}|p{6cm}|p{1.8cm}|p{4cm}|}
\caption{Image databases for COVID-19 research}
\label{table:databases}\\
\hline
Item & Dataset name & \# Images & Classes  \\
\hline
A    & HM Hospitales \citep{HMHospitals2020}                                             &   5,560  & COVID-19  \\
B    & BIMCV-COVID19 \citep{BIMCVDataPort}                                               &   3,013  & COVID-19  \\
C    & Actualmed COVID-19 chest X-rays \citep{GitHubInitiative}                          &     188  & COVID-19; Normal  \\
D    & ChinaSet - The ShenzhenSet  \citep{Jaeger2014TwoDiseases.}                        &     662  & Pneumonia; Normal \\
E    & Montgomery \citep{Jaeger2014TwoDiseases.}                                         &     138  & Pneumonia; Normal  \\
F    & ChestX-ray8  (CRX8)  \citep{Wang2017ChestX-ray8:Diseases}                         &  61,790  & Pneumonia; Normal  \\
G    & CheXpert   \citep{Irvin2019}                                                      &   4,623  & Pneumonia  \\
H    & MIMIC-CXR   \citep{Johnson2019a}                                                  &  16,399  & Pneumonia  \\
I    & COVID-19 Image Data Collection \citep{PaulCohen2020}                              &     481  & Viral (COVID-19; SARS; MERS; among others); Bacterial; Others     \\
J    & Kermany et al.    \citep{Kermany2018}                                             &   5,840  & Normal (1,575); Pneumonia Bacterial (2,771); non-COVID-19 lung infection (1,494)    \\
K    & RSNA, Radiopedia and SIRM    \citep{Dadario2020COVID-19Kaggle}                    &      73  & COVID-19    \\
L    & RYDLS-20  \citep{Pereira2020}                                                     &   1,144  & Normal (1,000); Pneumonia (144): (MERS (10); COVID-19 (90); Pneumocystis (11); SARS (11); Streptococcus (12); Varicella (10)) \\
M    & COVID-19 Radiography Database (Qatar university) \citep{Rahman2020COVID-19Kaggle} &  21,165  & COVID-19 (3,616); Normal (10,192); Viral Pneumonia (1,345); Non-COVID-19 lung infection (6,012)  \\
N    & NIH Chest X-ray  \citep{Wang2017ChestX-ray8:Diseases}                             & 108,948  & Atelectasis; Mass; Cardiomegaly; Nodule; Effusion; Normal; Infiltration; Pneumonia  \\
O    & Chest X-ray Images (Pneumonia)  \cite{Mooney2018ChestKaggle}                      &   5,863  & Pneumonia (bacterial and viral); Normal  \\
P    & COVID-19 dataset   \citep{2020BASESIRM}                                           &     115  & COVID-19   \\
Q    & CHUAC dataset  \citep{DeMoura2020}                                                &   1,616  & Normal (728); Pathological (648); COVID-19 (240)  \\
R    & COVID-19 X rays \citep{Dadario2020COVID-19Kaggle}                                 &      79  & COVID-19  \\    
\hline
\end{longtable}

\subsection{Approaches for the automatic detection of COVID-19 using X-ray images}

Table \ref{table:art_details} presents relevant information regarding the methodology applied in each selected paper, the best results achieved, the AI models used, the image databases involved, the classes, and the preprocessing operations.  
The state-of-the-art networks most used are the VGG and ResNet family and the preprocessing steps more prevalent are resizing, normalization and cropping. Most of the models reported have high performances, however, the experimental setups are not always clear due to most of the authors combine multiple databases and then split the resulting set in the training, validation, and test partitions without doing any further clarification. In many cases, the testing of the model is done using very few images and none of the analyzed works perform clinical validation of the methods.

\begin{longtable}[!ht]{|p{2.6cm}|p{2.2cm}|p{1.4cm}|p{2cm}|p{2.3cm}|p{2.2cm}|}
\caption{Implementation details and results of the selected papers for the detection of COVID-19 using chest X-ray images}
\label{table:art_details}\\
\hline
Article & Models  & Database  & Classes & Preprocessing   & Best Results   \\
\hline
 \citep{Apostolopoulos2020}  & VGG16; MobileNetV2; Inception; Xception; Inception-ResNet-V2
                             & I, J  
                             & COVID-19; Pneumonia; Normal 
                             & Resize, 200 x 266 & MobileNetV2:  \newline Accuracy, 94.7\%; \newline Sensitivity, 98.7\%; \newline Specificity, 96.5\% \\

\citep{Ucar2020}             & COVID Diagnosis-Net (based on SqueezeNet)                    
                             & I, O          
                             & COVID-19; Pneumonia; Normal                                                                                                                                                                                      
                             & Resize, 227 x 227; Shuffled; Normalization with the mean subtracting operation; Conversion to RGB with 8-bit depth & Accuracy, 98.3\%;\newline Specificity, 99.1\%; \newline F1, 98.3\%  \\

\citep{Ozturk2020}           & Dark CovidNet (based on Darknet-19 model)                    
                             & F, I          
                             & (COVID-19; Normal);  (COVID-19; Pneumonia; Normal)                                                                                                                                                                & N/A                                                                                                                
                             & Binary Accuracy, 98.1\%  \\

\citep{Togacar2020}          & MobileNetV2;	SqueezeNet; Social Mimic optimization method; SVM  
                             & I, M          
                             & COVID-19; Pneumonia; Normal
                             & Convertion to jpg format; Fuzzy Color technique                                                                    
                             & Overall accuracy, 98.3\%; \newline COVID-19 Sensitivity, 99.3\%; \newline COVID-19 Specificity, 99.4\%  \\

\citep{Pereira2020}          & SVM; MLP; DT; RF; Hierarchical Clus-HMC 
                             & I, N, L       
                             & Normal;	COVID-19; MERS; SARS; Varicella; Streptococcus; Pneumocystis                                                                                                                                             & Manual crop                                                                                                        
                             & Multiclass: \newline COVID-19 	class F1 score, 83.3\%  \newline Hierarchical: \newline COVID-19 class F1 score, 88.8\%   \\

\citep{Apostolopoulos2020b}  & MobileNetV2                                      
                             & I, P          
                             & COVID-19; Edema; Effusion; Emphys; Fibrosis; Pneumonia; Normal                                                                                                                                                    & Resize, 200 x 200                                                                                                  
                             & Accuracy between the seven classes of 87.7\%    \\

\citep{Waheed2020}           & ACGAN; VGG16                                               
                             & I, M, R       
                             & COVID-19; Normal                                                                                                                                                                                                  & Resize, 112 x 112 x 3; Normalization                                                                               
                             & Using actual data:  \newline	Sensitivity, 69.0\%; \newline Specificity, 95.0\%; \newline Accuracy, 85.0\%  \newline  Including synthetic images: \newline Sensitivity, 95.0\%; \newline Specificity, 90.0\%; \newline Accuracy, 97.0\% \\

\citep{Khan2020CoroNet:Images} & CoroNet                                                                                                                                     
                               & I, O          
                               & (COVID-19; Normal; Pneumonia bacterial; Pneumonia viral); (COVID-19; Pneumonia; Normal)                                                                                                                         & Resize, 224 x 224                                                                                                  
                               & Four classes: \newline Accuracy, 93.0\%; \newline Three classes: \newline Accuracy, 95.0\%    \\

\citep{Das2020}              & Truncated Inception Net   
                             & I, O, S       
                             & COVID-19; Pneumonia; Tuberculosis; Normal                                                                                                                                                                         & Resize, 224 x 224 x 3                                                                                              
                             & COVID-19 positive cases: \newline Accuracy, 99.96\%; \newline AUC, 100\%   \\

\citep{Toraman2020}          & CapsNet                
                             & I, N          
                             & (COVID-19; Normal); (COVID-19; Normal; Pneumonia)                                                                                                                                                                 & Resize, 128 x 128; Data 	augmentation                                                                              
                             & Binary class: \newline Accuracy, 97.2\%  \newline Multi-class: \newline Accuracy, 84.2\%   \\

\citep{Blain2021}            & U-Net; DenseNet121      
                             & K             
                             & N/A                                                                                                                                                                                                               & Lung segmentation                                                                                                  
                             & Diagnosing alveolar opacities:\newline Accuracy, 78.5\%  \newline	 Diagnosing interstitial opacities: \newline Accuracy, 90.7\%   \\

\citep{Horry2020COVID-19Data}  & VGG19      
                               & I, N          
                               & COVID-19; Normal; Pneumonia                                                                                                                                                                                     & Resize, 224 x 224; Histogram equalization using N-CLAHE                                                            
                               & X-ray: \newline Accuracy, 86.0\%  \newline  Ultrasound: \newline	Accuracy, 100\%  \newline  CT: \newline Accuracy, 84.0\%   \\

\citep{King2020}             & Self-Organizing Feature Map  
                             & I             
                             & COVID-19; Normal                                                                                                                                                                                                  & Resize                                                                                                             
                             & Euclidean distance of 1.1 between 1st and 2nd winning neurons  \\

\citep{Karar2020}            & VGG16; 	ResNet50V2; DenseNet169     
                             & I             
                             & Normal; COVID-19; Viral Pneumonia; Bacterial Pneumonia                                                                                                                                                            & Resize, 150 x 150                                                                                                  
                             & Accuracy, 99.9\%   \\

\citep{Ohata2021}            & MobileNet; DenseNet121; Inception-ResNet-V2; Bayes; RF; MLP; KNN; SVM                          
                             & I, R, O, N    
                             & COVID-19; Normal                                                                                                                                                                                                  & Resize, (224 x 224;  299 x 299; 331 x 331)                                                                         
                             & MobileNet + SVM (Linear): \newline Accuracy, 98.6\%; \newline F1-score, 98.5\%     \\

\citep{Shorfuzzaman2020}     & Siamese Network             
                             & I, O          
                             & COVID-19; Normal; Pneumonia                                                                                                                                                                                       & Resize; Normalization; Histogram-equalization                                                                      
                             & Accuracy, 95.6\%; \newline AUC, 98.9\% \\

\citep{DeMoura2020}          & DenseNet161      
                             & Q            
                             & COVID-19; Pathological; Normal; Combinations                                                                                                                                                                      & N/A                                                                                                                
                             & Accuracy, 90.3\% in (Normal \& Pathological) vs. COVID-19 \\

\citep{Nayak2021}            & AlexNet; VGG16; GoogleNet; MobileNetV2; SqueezeNet; ResNet34; ResNet50; InceptionV3     
                             & F, I          
                             & COVID-19; Normal                                                                                                                                                                                                  & Normalization                                                                                                      
                             & ResNet34, 98.3\%   \\

\citep{Albahli2021}          & CNN Designed   
                             & I, O          
                             & (COVID-19; Normal); (COVID-19; Normal; Bacterial Pneumonia)                                                                                                                                                       & Resize, 224 x 224                                                                                                  
                             & Binary classification: \newline Accuracy, 98.7\%; \newline Sensitivity, 100\%; \newline Specificity, 98.3\%    \\

\citep{Albahli2021Using}     & NasNetLarge; Xception; InceptionV3; Inception-ResNetV2; ResNet50                                                                            
                             & I, N          
                             & COVID-19; normal; 14 other chest diseases                                                                                                                                                                         & Histogram equalization; Lung and heart segmentation                                                                
                             & First classifier: \newline Accuracy, 96.3\%  \newline   Second classifier: \newline Accuracy, 87.8\%     \\

\citep{Singh2021a}           & VGG19; VGG16; ResNet50; DenseNet161; DenseNet169; Naive Bayes   
                             & Q, R, M, N, G 
                             & COVID-19; Pneumonia; Normal                                                                                                                                                                                       & Histogram equalization (CLAHE); Dynamic image filtering (NLMD)                                                     
                             & Accuracy, 98.7\%   \\

\citep{Sheykhivand2021}      & GANs; LSTM networks  
                             & O, R, I, M    
                             & (Normal; COVID-19);  (Normal; Pneumonia (Viral; Bacterial; COVID19)); (Normal; COVID-19; Pneumonia Viral and Pneumonia Bacterial); (Normal; COVID-19; Pneumonia Bacterial); (Normal; COVID-19; Pneumonia Viral); (COVID-19; Pneumonia Bacterial; Pneumonia Viral); & Resize, 224 x 224; Normalization                                                                                   
                             & Two classes: Accuracy, 99.5\%     \\

\citep{Tuncer2021}           & SVM   
                             & I, O          
                             & COVID-19; Pneumonia; Normal   
                             & Fuzzy transform   
                             & Accuracy, 97.0\% \\
\hline
\end{longtable}

We summarize the different approaches in the literature for COVID-19 automatic detection using X-ray images. We aim to present the different strategies for preprocessing, classification, and interpretability implemented in literature.

\subsubsection{Preprocessing strategies} 
The most common preprocessing methods found are: normalization, cropping and resizing, being the predominant size 224 x 224 \citep{SarvAhrabi2021}. It has also been applied the selection of only anteroposterior (AP) or posteroanterior (PA) views from the images in the databases \citep{Arias-Londono2020ArtificialApproach, Panwar2020ApplicationNCOVnet} and the state-of-the-art architectures own preprocessing \citep{Castiglioni2021}. Other strategies found include the Fuzzy Color technique \citep{Togacar2020} and the multiscale offline augmentation, which incorporate shearing the image, adding Gaussian noise, and decreasing the brightness \citep{Ucar2020}.

Authors have stated that the CNNs used for the detection of COVID-19 using chest X-ray images, base their classifications in areas in the input image outside the region of interest, which has no relation with COVID-19 signs \citep{Majeed2020}. Therefore, some works perform a segmentation of the lungs using the U-Net network as a preprocessing step to ensure that the network classification is based on regions inside them \citep{Tartaglione2020UnveilingData, Arias-Londono2020ArtificialApproach, Rajaraman2020}; another approach is presented by \citep{Aslan2021} here the lungs are segmented before the classification using a new ANN-based network.

Several authors include data augmentation in the network training to mitigate the dataset imbalance, the most common transformations reported are rotation, horizontal, and vertical flip \citep{Khan2020CoroNet:Images, Zebin2021}. Other transformations applied are: Gaussian noise \citep{Nayak2021}; shearing, elastic distortion \citep{SarvAhrabi2021} and histogram equalization \citep{Tartaglione2020UnveilingData}. Data imbalance has also been compensated using Generative Adversary Networks (GANs) to generate artificial images and augment the minority COVID-19 class, some examples are the CycleGAN \citep{Albahli2021}; the conditional generative adversarial networks \citep{Zebin2021}, and the Auxiliary Classifier Generative Adversarial Network (ACGAN) \citep{Waheed2020}, this work shows that their model accuracy can increase by 10\% if the synthetic images produced are used during training.

\subsubsection{Classification methods}
Despite the short time since COVID-19 emerged, the limited availability of data and knowledge in this regard, many methods have been proposed for the automatic detection of this disease using chest X-ray images.
    
Transfer learning using state-of-the-art CNNs architectures is the most commonly found strategy to perform this task. Among the most frequently used networks are VGG19, VGG16, Inception-ResNetV2, MobileNetV2, ResNet50, and EfficientNetB0 \citep{Apostolopoulos2020, Zebin2021}. 

\begin{figure}[!ht]
    \centering
    \includegraphics[width=\textwidth]{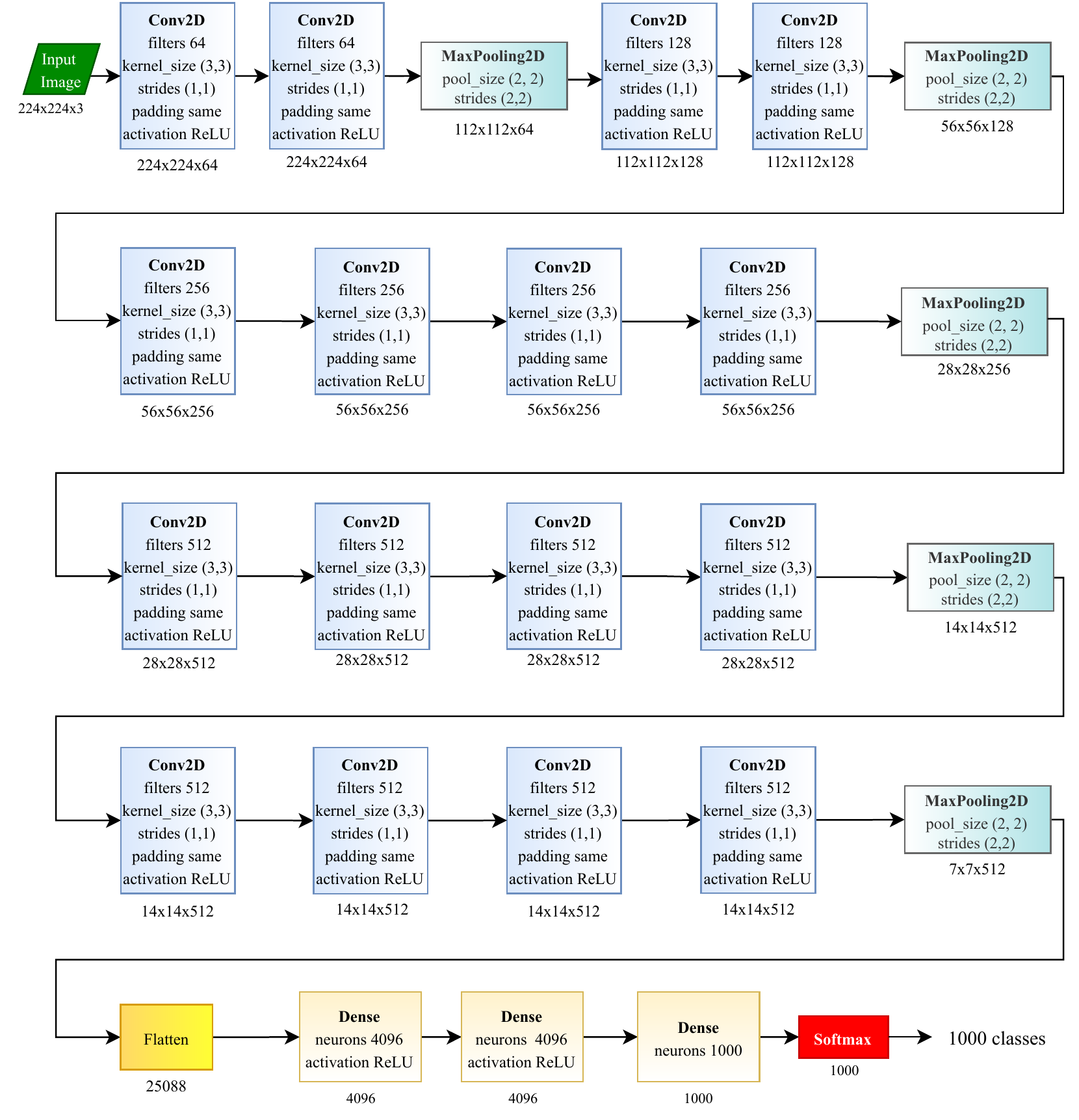}
    \caption{Original architecture of the VGG19 convolutional neural network.}
    \label{fig:vgg}
\end{figure}

Figure \ref{fig:vgg} shows the VGG19 architecture that has a total of 19 layers, namely 16 2D-Convolutional layers and 3 dense layers. This figure shows the number of filters, the kernel size, the strides size, the padding name, the activation function, and the shape of the output of each layer. Usually, to implement transfer learning, researchers deleted the layers after the Flatten, i.e., the dense layers and Softmax activation function, and replace them with a particular fully connected layer. The final activation function generates predictions from the last dense layer, with a specific number of classes. 

Other state-of-the-art CNNs architectures have also been used in literature: DenseNet161 is adapted to classify chest X-ray images acquired by portable equipment \citep{DeMoura2020}; ResNet50 and ResNet101 are used in a two-stage model, where pneumonia and healthy images are initially identified and later COVID-19 examples are classified from the pneumonia cases \citep{Jain2020AImages.}; ResNet50 lead to the best results of a two stages model to classify COVID-19 and other 14 chest diseases, being the results competitive with currently used state-of-the-art models \citep{Albahli2021Using}; Inception is truncated at a layer that is chosen experimentally to avoid possible overfitting due to the lack of COVID-19 positive samples \citep{Das2020}. Similarly, ImageNet pre-trained models are also pruned and their predictions are combined through different ensemble strategies \citep{Rajaraman2020}.
    
A common approach uses state-of-the-art CNNs as feature extractors and machine learning classifiers to make the final predictions, this integrated approach has been performed using machine learning models like Support Vector Machine (SVM), linear kernel and radial basis function, k-nearest neighbor, Decision Tree, CN 2 rule inducer techniques and deep learning models like MobileNetV2, ResNet50, GoogleNet, DarkNet, and Xception \citep{Ohata2021, Mohammed2021}. Additional models are also used, namely; MobileNetV2 and SqueezeNet are used to extract characteristics that are later processed using the Social Mimic optimization method and the final classification task is performed by a SVM \citep{Togacar2020}. A similar approach is made by \cite{Ismael2021DeepImages.} using VGG16, VGG19, ResNet18, ResNet50, and ResNet101 and SVM. In \citep{Sahlol2020} Inception is used to extract features, and Marine Predators Algorithm, a swarm‑based feature selection algorithm is used to select the most relevant features; a fuzzy tree transformation is applied to each chest image and then exemplar division, a novel machine learning model is used. Then features are obtained using a multi-kernel local binary pattern, the most important are selected using the iterative neighborhood component and finally, conventional classifiers perform the classification. The best performance is obtained using a cubic SVM \citep{Tuncer2021}.

Implementations of traditional machine learning methods and strategies like late fusion, early fusion, and hierarchical classification are used to classify not only COVID-19 but also up to 14 other lung diseases \citep{Pereira2020, Yoo2020DeepImaging.}. 
    
Other methodologies were also found, some of the most representative are: an ensemble of ten convolutional neural networks based on ResNet50 architecture and ML models \citep{Singh2021a, Castiglioni2021}; a cascaded classification scheme using pre-trained CNN architectures \citep{Karar2020}; a multi-kernel CNN block combined with pre-trained ResNet34 to overcome imbalance in the dataset \citep{Mursalim2021}; an integration of contrastive learning with a fine-tuned pre-trained ConvNet encoder to capture unbiased feature representations and a Siamese network, which makes the final classification \citep{Shorfuzzaman2020}; an unsupervised network called Self-Organizing Feature Maps, which is analyzed using the saliency of features, the authors state that the unsupervised method can extract features that allow to accurately classify the COVID-19 chest X-ray images \citep{King2020}; a multimodal approach using clinical and radiographic features, both are compared using the unpaired student’s t-test or Mann-Whitney U test and the segmentation and detection of opacities are also carried out \citep{Blain2021}.
    
Many authors also implement their own networks, some of them are based on state-of-the-art CNNs, such as VGG16 \citep{Panwar2020ApplicationNCOVnet}, AlexNet \citep{Aslan2021} and Xception \citep{NarayanDas2020AutomatedX-rays}. In \citep{Ozturk2020} a network based on DarkNet is evaluated by a radiologist, who concludes that the model has a good performance detecting COVID-19 cases for the binary class task, but it makes incorrect predictions in poor quality chest X-ray images and patients with acute respiratory distress syndrome. The CoroNet network is proposed based on the Xception architecture pre-trained on the ImageNet dataset. It is trained using X-ray images of COVID-19 and other pneumonia, the obtained network weights are publically available \citep{Khan2020CoroNet:Images}. COVIDiagnosis-Net includes fine-tuning of the SqueezeNet using a bayesian optimization method and offline augmentation only to COVID-19 class \citep{Ucar2020}.
    
 Some authors propose their architectures from scratch, \cite{SarvAhrabi2021} propose a network with 12 layers including convolution, max-pooling, batch normalization, dropout, activation, and fully-connected layers; \cite{Hussain2021} present a 22-layer CNN model, that is evaluated by a clinician in multiple classification scenarios, 2 classes, 3 classes, and 4 classes.
     
Most of the found studies conduct a binary classification, COVID-19 vs Normal cases. However, other works attempt multiclass classification, some of them include: COVID-19, pneumonia and no-findings \citep{Ucar2020, Toraman2020}; COVID-19, pneumonia viral, pneumonia bacterial and no-findings \citep{Khan2020CoroNet:Images}, \citep{Jain2020AImages.}; COVID-19, tuberculosis and non-findings \citep{Das2020, Yoo2020DeepImaging.}, and COVID-19 severity classification \citep{Blain2021}. Some authors explore multiple combinations of those scenarios \citep{Hussain2021, Sheykhivand2021, Majeed2020}, and others take advantage of the availability of other lung-related decease labels in the datasets and perform the classification of COVID-19 examples along with multiple other pulmonary diseases \citep{Apostolopoulos2020b, Albahli2021Using}.

\subsubsection{Interpretability and CNN benchmarking} 
\cite{Nayak2021}, analyze the performance of eight state-of-the-art CNNs, they tune the number of trainable layers of the network, nodes, epochs, layers of the classifier placed at the top of the network, the batch size, the learning rate, and the optimizer algorithm, they find that the ResNet family of architectures has the highest classification accuracy; a similar analysis is implemented by \citep{Majeed2020} using transfer learning and 12 state-of-the-art CNNs architectures, a critical analysis that includes the needed time to train each network is also presented, in the binary scenario the best results were obtained using the networks Xception, Inception-ResnetV2, and SqueezeNet; likewise, \cite{Apostolopoulos2020} implement transfer learning to compare 5 state-of-the-art CNNs, the best result are obtained using the VGG19. 

\cite{Pham2021ClassificationTuning}, compare the performance of state-of-the-art fine-tuned CNNs and recently developed networks like CovidGAN, CoroNet, and DarkCovidNet, using an experimental setup as similar as possible to the original studies. Despite exact comparisons are not possible due to databases updates, it is concluded that similar performances are obtained using relatively small CNNs like AlexNet or SqueeNet and the new and sometimes more complex architectures. \cite{Tartaglione2020UnveilingData}, provide insights and also raise warnings regarding the generality of the results of COVID-19 classification using deep learning and chest X-ray images. Alternately \cite{Horry2020COVID-19Data} compare the results achieved training the VGG16 network using CT, X-ray, and Ultrasound chest images to classify COVID-19, pneumonia and healthy subjects. The best result is achieved using Ultrasound images.
    
One of the main drawbacks of deep learning is the lack of interpretability, which has imitated its application in some areas \citep{Bassi2020a}. That is why, some authors have included visualization methods that could provide credibility and increase trust in users. Particular focus has been given to heat maps. In \citep{Ozturk2020} is proposed a network that provides a heat map along with the classifications, results are evaluated by a radiologist who concludes that `The heat map showed a greater concentration area in the X-rays of patients with COVID-19 than the area in which the disease is not seen'. Besides heat maps, t-SNE is also used to improve explainability \citep{Arias-Londono2020ArtificialApproach}, as well as class activation maps \citep{Ucar2020}.
    
Finally, a widely used and accepted visualization method is the grand cam, which is used in multiple works \citep{Singh2021a, Liang2021}.

\subsection{Metrics used in the evaluation of algorithms}
Most of the metrics defined below are expressed for binary classification tasks in terms of the numbers of True Positive (TP) predictions, True Negatives (TN), False Positives (FP), and False Negatives (FN). TP refers to the positive instances correctly classified as positive; TN refers to the negative instances correctly classified as negative; FP refers to the negative instances incorrectly classified as positive; FN refers to the positive instances incorrectly classified as negative.
\begin{itemize}
    \item Accuracy: proportion of correct predictions, usually presented as a percentage or as a number from 0 to 1.
    \begin{equation}
        Accuracy = \frac{TP+TN}{TP+TN+FP+FN}
    \end{equation}
    
    \item Specificity: it measures the ability of the classifier to correctly identify the negative instances of a class.
    \begin{equation}
        Specificity=\frac{TN}{TN+FP}
    \end{equation}
    
    \item Recall/Sensitivity: also known as recall, it measures a classifier's ability to correctly identify the positive instances of a class.
    \begin{equation}
        Recall=\frac{TP}{TP+FN}
    \end{equation}
    
    \item Precision/Positive Predictive Value: also known as precision, it represents the proportion of positive cases among the instances classified as positive.
\begin{equation}
    Precision = \frac{TP}{TP+FP}
\end{equation}
    
    \item F1-score: weighted harmonic mean of specificity and sensitivity normalized between 0 and 1. This metric considers imbalanced classes, and it is useful when a task requires high specificity and sensitivity.
    \begin{equation}
    F1 = 2\frac{Precision~ Recall}{Precision+Recall}
    \end{equation}
    
    \item Kappa statistics (K): standard measure of agreement between the expected and the observed number of correct predictions. The expected number of correct predictions is computed from class probabilities, making it suitable for evaluating multiple classes in imbalanced datasets. The equation represents the expected number of agreements as Pe and the observed as Po.
    \begin{equation}
        K = \frac{Po - Pe}{1-Pe}
    \end{equation}
\end{itemize}

\section{Conclusions and future work} 
Given the worldwide impact of the COVID-19 pandemic, it is a priority to develop alternative diagnostic tools available for everyone and offer reliable results. This paper presents a systematic literature review of AI applied to COVID-19 diagnosis using chest X-ray imaging. It includes the different preprocessing techniques, classification methods using ML algorithms, strategies to increase the interpretability of the models, and the articles that perform a critical analysis of the state-of-the-art and the new architecture designed to perform this task. 

The main limitation researchers have faced when developing these systems is the quality and availability of data. To overcome this situation, the use of preprocessing techniques, such as histogram equalization, lung segmentation, data augmentation using rotation or cropping operations, and synthetic data generation using GANs, have been implemented to improve the detection performances of the models. The most frequent approach for the classification of COVID-19 from X-ray images is transfer learning using pre-trained CNNs architectures, such as VGG16, DenseNet121, and ResNet50. Other proposals involve state-of-the-art architectures as feature extractors and traditional ML methods as classifiers. There are also multiple novel CNNs explicitly designed for this task which allows more flexibility and potentially smaller networks by narrowing the scope of the classification task. Overall, literature methods have exceptional results with classification accuracies over 95\% and even 98\%, however, the test set and the quality of the data, are usually unclear.

Regarding the questions set at the start of the paper, we have shown that (1) conventional image classification CNNs with pre-trained weights using ImageNet, or more complex approaches as Capsule or Siamese networks have been used to diagnose COVID-19 from chest X-ray images; (2) current detection percentages are over 98\% accuracy in binary classification (COVID-19 and Normal). However, no clinical trials have been performed in none of these models and the experimental setups are usually unclear; (3) this paper compiles an active set of databases for training and evaluating AI models, despite the relatively high number of available databases of chest X-rays, there is a limited amount of labeled COVID-19 cases, which leads researchers to combine various databases; and in (4) most of the papers that use lung segmentation as a preprocessing step, do so using a U-Net architecture.
    
Based on the present literature review, we identify possible research opportunities as follows:
\begin{itemize}
    \item Construct or contribute to databases of chest X-ray images aiming to create a representation of the different characteristics of real-world images, allowing proper benchmarking and future model proposals.
    \item Develop new CNNs for image segmentation, focusing on the segmentation of lungs and radiological findings associated with COVID-19 disease.
    \item Broaden the classification scope to detect factors such as severity, or disease progression.
    \item Design new preprocessing operations or pipelines, taking into account, for example, the removal of artifacts or medical devices such as necklaces, tubes, or ECG lead wires.
    \item Detect COVID-19, and its outcome considering other clinical variables such as the patient's history.
    \item Perform transfer learning based on networks that have been trained for other lung diseases.
    \item Design architectures or computational elements of CNNs for the detection of COVID-19 from validated reference models.
\end{itemize}

\section*{Acknowledments}
The authors acknowledge to SES Hospital Universitario de Caldas, Alcaldía de Manizales, the Universidad Autónoma de Manizales (UAM) and Minciencias from Colombia, and the Mixed Unit of Biomedical Imaging FISABIO-CIPF from Spain, for their contributions and funding through the project “Detección de COVID-19 en imágenes de rayos X usando redes neuronales convolucionales” with code 699-106 from UAM and contract 831 from grant 874-2020 of Minciencias.  

\section*{Disclosure Statement}
No potential conflict of interest was reported by the authors.

\section*{Appendix}

Table \ref{tab:acronyms} shows some acronyms and their meanings.

\begin{table}[!ht]
\centering
\caption{Acronyms}
\label{tab:acronyms}
\begin{tabular}{ll}
\hline
\textbf{Acronym} & \textbf{Description}                                      \\
\hline
ACGAN            & Auxiliary Classifier Generative Adversarial Network       \\
AI               & Artificial Intelligence                                   \\
CNNs             & Convolutional Neural Networks                             \\
CT               & Computer tomography                                       \\
DL               & Deep Learning                                             \\
FN               & False Negatives                                           \\
FP               & False Positives                                           \\
GANs             & Generative Adversarial Network                            \\
K                & Kappa statistics                                          \\
ML               & Machine Learning                                          \\
RT-PCR           & Real-time reverse transcription polymerase chain reaction \\
SARS-CoV-2       & Severe Acute Respiratory Syndrome coronavirus 2           \\
SVM              & Support Vector Machine                                    \\
TN               & True Negatives                                            \\
TP               & True Positive                                             \\
WHO              & World Health Organization                                 \\
\hline
\end{tabular}
\end{table}

\bibliographystyle{unsrtnat}







\end{document}